\documentclass[11pt,aps,prd,showpacs,amsmath,amssymb,floatfix,nofootinbib,superscriptaddress]{revtex4-2}

\usepackage{graphicx}  
\newcounter{subfig}

\usepackage{float} 
\usepackage{multirow}
\usepackage[dvipsnames]{xcolor}
\usepackage[colorlinks=true,allcolors=BlueViolet]{hyperref}
\usepackage{multirow,bigdelim}
\usepackage[dvipsnames]{xcolor}

\usepackage{latexsym}
\usepackage{amsfonts}
\usepackage{amsmath,amssymb}
\usepackage{slashed}
\usepackage{color}
\usepackage{orcidlink}
\usepackage{supertabular} 
\usepackage{epsfig}
\usepackage{lipsum}

\newcommand{\Tr}{\text{Tr}}

\newcommand{\ific}{\affiliation{Instituto de F\'isica Corpuscular (centro mixto CSIC-UV), \\
Institutos de Investigaci\'on de Paterna, Apartado 22085, 46071, Valencia, Spain}}

\newcommand{\uv}
{\affiliation{Departamento de F\'{\i}sica Te\'orica and IFIC, Centro Mixto Universidad de Valencia-CSIC Institutos de Investigaci\'on de Paterna, Aptdo.22085, 46071 Valencia, Spain}
}

\begin{document}

\title{The $B^{(*)}\bar{K}^{(*)}$-coupled-channel system in the hidden-gauge approach}

\author{J. Sánchez-Illana}
\email{Jesus.Sanchez@ific.uv.es}
\uv

\author{R. Molina}
\email{Raquel.Molina@ific.uv.es}
\uv

\author{Pan-Pan~Shi} \email{Panpan.Shi@ific.uv.es}
\uv\ific

\begin{abstract}
In this work we provide predictions for bottom-strange molecular states within the Hidden Gauge Formalism. We study the coupled-channel scattering of $B^{(*)}\bar{K}^{(*)}$ states and, by fixing only one free parameter to obtain the mass of a new excited $B_s^0$ state seen by the LHCb, we predict the pole parameters of six states in this sector. Concretely, we get that the masses of the flavor partners of the $D_{s0}(2317)$ and $D_{s1}(2460)$ states in the bottom sector are $5760$ and $5802$ MeV for the $B\bar{K}$ ($J^P=0^+$) and $B^{*}\bar{K}$ ($1^+$) states, respectively. Moreover, the recently seen states by the LHCb with masses around $6100$ and $6160$ MeV can be interpreted as $B\bar{K}^*$ and $B^*\bar{K}^*$ molecular states, according to reasonable values of the pole parameters and the splitting between these two states obtained in our calculation.
\end{abstract}

\maketitle

\section{Introduction}
According to the Heavy Quark Symmetry (HQS), the interaction of a heavy quark with a light one becomes independent on the spin and flavor of it in the heavy quark limit~\cite{Isgur:1989vq,Isgur:1990yhj,Isgur:1991wq,Manohar:2000dt}. This implies that one can arrange the states into doublets according to $j_{\bar{q}}^P$ (the full spin of the light degrees of freedom). Then, for $j_{\bar{q}}^P=\frac{1}{2}^+$, one has a $J^P=0^+,1^+$ doublet, while for $j_{\bar{q}}^P=\frac{3}{2}^+$ one gets another doublet with spin-parity $J^P=1^+,2^+$. As is well-known, in the charm sector these $c\bar{s}$ states would in principle correspond to the $D_{s0}(2317)$ and the $D_{s1}(2460)$ for the first doublet and to the $D_{s1}(2536)$ and the $D_{s2}(2573)$ in the second case. The fact that the observed states in the first one are close to the $DK$ and $D^*K$ thresholds and are much narrower than their quark model predictions has led to a growing acceptance of the molecular description for these states~\cite{barnesclose,Kolomeitsev:2003ac,faesslergutsche,faesslergutsche2,Gamermann:2006nm,Gamermann:2007fi}. This is also supported by Lattice QCD simulations (LQCD), since $DK$($D^*K$) interpolators are needed to extract the masses of these states~\cite{bali,mohler,lang,alberto,cheungthomas}, and the analysis of recent lattice data through Effective Field Theory methods, as those employing unitary scattering amplitudes based on the Hidden Gauge Formalism (HGF) or Heavy Meson Chiral Perturbation Theory (HMChPT) ~\cite{alberto,liuorginos,pedro,menglin,Gil-Dominguez:2023puj}. These precise analyses, which cover most LQCD simulations, conclude that around $70\%$ of the states in the $J^P=0^+,1^+$ doublet is of molecular nature, while being the $c\bar{s}$ seed also present. Heavy Flavour Symmetry (HFS) suggests that if those $D_s^*$ states exist as predominantly two-meson molecules, similar states should also be present in the bottom sector~\cite{Manohar:2000dt,Guo:2013sya,Guo:2017jvc}. Thus, it is reasonable to ask about  predictions from EFT methods in this sector. In this article, we will focus on possible predictions of the heavy flavor partners of the $D_{s0}(2317)$ and $D_{s1}(2460)$ states, and their spin partners in the bottom sectors from the Hidden Gauge formalism (HGF). 

Some previous calculations in this regard are made in the literature. First of all, the $P$-wave states in the quark model have masses around $5805-5880$ MeV~\cite{Godfrey:1986wj,Ebert:1997nk,DiPierro:2001dwf,Ebert:2009ua,Sun:2014wea,Godfrey:2016nwn}. Three states have been observed in this region, two of them that are close to the expectations of the quark model for the narrow $1^3P_2$ and $1P_1$ states with $J^P=1^+, 2^+$, and masses around $30$ MeV lower than the experimental ones. However, the state with $J^P=0^+$, not yet been observed, has radically different properties in the quark model prediction from the expectation of the bottom partner of the $D_{s0}(2317)$ in unitarized EFT's. In the latter case it should be a very narrow state close to the $B\bar{K}$ threshold, while in the quark model, the $1^3P_0$ state has a mass near $5830$ MeV and a width around $140$ MeV~\cite{Godfrey:2016nwn}. Although the $0^+$ state has not been discovered, there are some predictions from LQCD~\cite{Gregory:2010gm,Lang:2015hza,Ni:2023lvx,Gayer:2024akw}. These simulations indicate the existence of a bound state below the $B\bar{K}$($B^*\bar{K}$) threshold with spin-parity $J^P=0^+(1^+)$ and a binding energy of a few tens of MeV's. See also~\cite{Yang:2022vdb} for the investigation of the $B_s$ excited spectrum combining quark model, coupled-channel effects and LQCD. These states found in LQCD have masses and decay widths that differ from the quark model predictions. The energy levels evaluated in~\cite{Lang:2015hza} are analyzed by the authors of ~\cite{Albaladejo:2016ztm} in the framework of HMChPT, 
 concluding that the masses of the bottom partners, $B^*_{s0}$ and  $B^*_{s1}$, are $5710$ and $5760$, respectively. See also other two-meson molecular predictions in unitarized chiral EFT approaches with heavy quark symmetry~\cite{Kolomeitsev:2003ac,Bardeen:2003kt,Guo:2006rp,Guo:2006fu,Colangelo:2012xi,Altenbuchinger:2013vwa,Sun:2018zqs}. 

 In this article, we would like to dig into the $B_s^*$ spectrum by a different approach based on the known experimental data in this sector. In the PDG, two $B_s^*$ states with masses $6063.5\pm1.2$ ($6108.8\pm 1.1$) and $6114\pm 3$ ($6158\pm 4$) MeV have been observed with decay widths $\Gamma=26\pm 4$ ($22\pm 5$) and $66\pm 18$ ($72\pm 18$), respectively, that can decay into $B\bar{K}$($B^*\bar{K})$, see~\cite{LHCb:2020pet}. Although their quantum numbers have not yet been experimentally determined, both states can decay to $B^{\pm}K^{\mp}$ or $B^{*\pm}K^{\mp}$ with a missing photon from decay $B^*\to B\gamma$. This observation suggests that their quantum numbers could be $J^P=1^{+}$.
  In the quark model of~\cite{Godfrey:2016nwn}, there are only two $D$-wave states, the $1D_2$ and $1^3D_3$, with masses $6098$ and $6109$~MeV, respectively, that are predicted to be narrow states with decay widths close to $20$~MeV,\footnote{This is in the quark model called ARM in~\cite{Godfrey:2016nwn}. In the Godfrey-Isgur relativized quark model (GI) these states have masses $6169$ and $6179$ MeV for the  $1D_2$, $1^3D_3$ states, respectively. Therefore, the splitting is essentially $10$ MeV.} because the other predictions have widths near $200$ MeV. Note that the splitting of these states is of $11$ MeV, smaller than the observed one, of $53$ MeV. In addition, the quantum numbers have not been measured yet. Thus, here we will explore the possibility that these states could correspond instead to two-meson molecular candidates, and by assuming that the state in~\cite{LHCb:2020pet} with a mass of $6109$ MeV does correspond to a $B\bar{K}^*$ molecular state within the HGF, we will make predictions on the properties, as the masses and decay widths of the $B^{(*)}\bar{K}^{(*)}$ molecular states, with quantum numbers $J^P=0^+, 1^+$ and $2^+$. 

This paper is organized as follows. In Sect.~\ref{Sec:formula}, we construct the effective potentials of the isoscalar $B^{(*)} \bar K^{(*)}$ systems and the three-body dynamics. The pole positions and the relevant effective couplings in the $B^{(*)} \bar K^{(*)}$ systems are presented in Sect.~\ref{sec:result}. A summary is given in Sect.~\ref{sec:sum}. The loop function, including the self-energy contributions, is derived in Appendix~\ref{sect:kw}. In Appendix~\ref{Sec:partial_wave}, we provide the details for the partial-wave decomposition.

\section{Formalism}\label{Sec:formula}

 In this work we consider the system $B{}^{(*)}\bar K{}^{(*)}$. The channels $ B\bar K{}^{*}$, $ B^*\bar K$ and $ B{}^{*}\bar K{}^{*}$ can be connected through anomalous $VVP$ couplings by exchanging a pseudoscalar or vector meson. Since these are negligible close to the relevant threshold energies, we neglect them here. The $B\bar{K}$ and $ B{}^{*}\bar K{}^{*}$ channels are separated in energy by around $450$ MeV. {However, the coupling between these channels can lead to decay width of the states and is not driven by anomalous couplings. Hence, we consider two approaches, the on-shell approach based on the Bethe-Salpeter equation with a single channel, and the off-shell approach by solving the Lippmann-Schwinger equation with the coupled-channels systems:  1) $B\bar{K}$}-$B^*\bar{K}^*$ ($J^P=0^+,2^+$); 2) $B^*\bar{K}-B\bar{K}^*$ ($J^P=1^+$); and 3) $B^*\bar{K}^*$ ($J^P=1^+$). Light-(pseudoscalar, vector) meson exchanges are considered between these channels ($\pi$, $\eta$, $\eta^\prime$, $\rho$, $\omega$). The interaction is evaluated within the HGF. The Lagrangian for the $3V$ and $PPV$ vertices are~\cite{Nagahiro:2008cv,Molina:2009ct,Molina:2010tx},
\begin{align}
\mathcal{L}_{I I I}^{(3 V)}=i g\Tr\left[\left(\partial_\mu V_\nu-\partial_\nu V_\mu\right) V^\mu V^\nu\right];\quad 
\mathcal{L}^{PPV}=i g \operatorname{Tr}\left(\left[\partial_\mu \phi, \phi\right] V^\mu\right),
\label{Eq:Lagrangian_SU4}
\end{align}
and the coupling constant is given by $ g=m_{\rho}/2f_{\pi}$. The matrices for the pseudoscalar $P$ and vector  $V_{\mu}$ mesons are given by,
\begin{align}\label{eq:p}
P=\left(\begin{array}{cccc}
\frac{\pi^0}{\sqrt{2}}+\frac{\eta}{\sqrt{3}}+\frac{\eta^{\prime}}{\sqrt{6}} & \pi^{+} & K^{+} & B^+ \\
\pi^{-} & -\frac{\pi^0}{\sqrt{2}}+\frac{\eta}{\sqrt{3}}+\frac{\eta^{\prime}}{\sqrt{6}} & K^0 & B^0 \\
K^{-} & \bar{K}^0 & -\frac{\eta}{\sqrt{3}}+\sqrt{\frac{2}{3}} \eta^{\prime} & B_s^0 \\
B^- & \bar B^{0} & \bar B_s^{0} & \eta_b
\end{array}\right).
\end{align}
and
\begin{align}\label{eq:v}
V_\mu=\left(\begin{array}{cccc}
\frac{\rho^0}{\sqrt{2}}+\frac{\omega}{\sqrt{2}} & \rho^{+} & K^{*+} & B^{* +} \\
\rho^{-} & -\frac{\rho^0}{\sqrt{2}}+\frac{\omega}{\sqrt{2}} & K^{* 0} & B^{*0} \\
K^{*-} & \bar{K}^{* 0} & \phi & B_s^{*0} \\
B^{* -} & \bar B^{*0} & \bar B_s^{*0} & \Upsilon
\end{array}\right)_\mu, 
\end{align}
respectively. The relevant Feynman diagrams are depicted in Fig.~\ref{Fig:diagram}. Note that the use of SU(4) here is merely formal and the interaction is broken to SU(3). These lagrangians are shown to respect Heavy-Quark-Symmetry (HQS) for light meson exchanges between the interaction of heavy and light mesons since the heavy quark is acting as an spectator~\cite{Xiao:2013yca}. 
In Table~\ref{Tab:pot2}, we provide the result for the interaction kernel $V$ given by the above Lagrangians. Other possible meson exchanges, such as the $\sigma$-meson or heavy meson exchange, are not included here because these are negligible~\cite{Aceti:2014kja,Aceti:2014uea,Dias:2014pva}. 
\begin{figure}[tbh]
    \includegraphics[height=0.25\columnwidth]{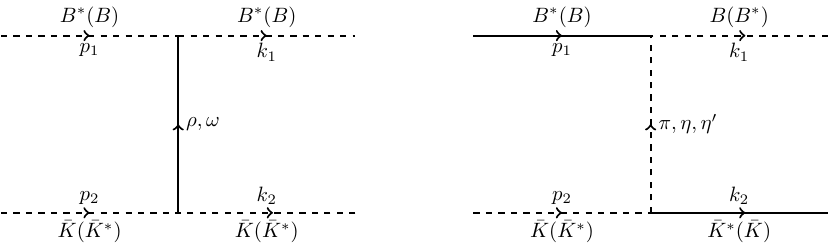}
   \caption{Feynman diagrams for the interaction of the $B^\ast \bar K-B\bar K^\ast$  systems.}
    \label{Fig:diagram}
\end{figure}
\begin{figure}[tbh]
    \includegraphics[height=0.25\columnwidth]{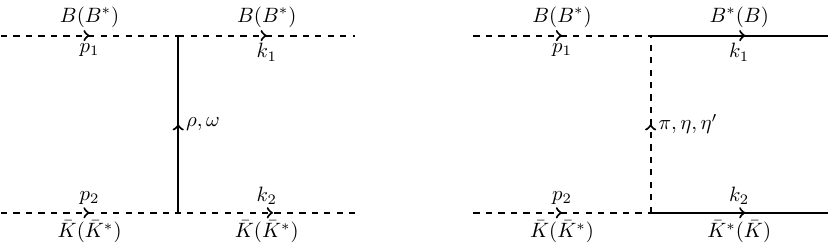}
   \caption{Feynman diagrams for the interaction of the $B \bar K-B^\ast\bar K^\ast$  systems.}
    \label{Fig:diagram2}
\end{figure}

\begin{table}[H]
\caption{{Interaction for the different coupled-channel systems considered contributing to the quantum numbers $J^P$. The partial-wave amplitude and, in particular, the $S$-wave projection for the different $J^P=0^+,1^+,2^+$ are given in Eqs.~(\ref{pwaveeq}), (\ref{Eq:epsilon_p1_k1}) and (\ref{Eq:epsilon_p1_p2}). }}
\label{Tab:pot2}
\renewcommand{\arraystretch}{1.2}
\begin{tabular*}{\columnwidth}{@{\extracolsep\fill} l | l}
\hline\hline 
    Channel ($J^P)$ &  \begin{minipage}[c]{0.72\columnwidth}\centering Potential \end{minipage}  \\[6pt] 
\hline
     $ B\bar K-B^\ast \bar K^\ast$ ($0^+$, $2^+$)
   &
     \begin{minipage}[c]{0.72\columnwidth}\centering
     \noindent\\[4pt]
     $V_{11}=\displaystyle\frac{g^2}{2}\bigl(3D_{su}(m_\rho)+D_{su}(m_\omega)\bigr)$\\[4pt]
     $V_{12}=\displaystyle\frac{2g^2}{3}\bigl(-4D_t(m_\eta)+D_t(m_{\eta^\prime})-9D_\pi\bigr)\,p_1\!\cdot\!\varepsilon^\ast(k_1)\;p_2\!\cdot\!\varepsilon^\ast(k_2)$\\[4pt]
     $V_{21}=\displaystyle\frac{2g^2}{3}\bigl(-4D_t(m_\eta)+D_t(m_{\eta^\prime})-9D_\pi\bigr)\;k_1\!\cdot\!\varepsilon(p_1)\,k_2\!\cdot\!\varepsilon(p_2)$\\[4pt]
     $V_{22}=\displaystyle\frac{g^2}{2}\bigl(3D_{su}(m_\rho)+D_{su}(m_\omega)\bigr)\,\varepsilon(p_1)\!\cdot\!\varepsilon^\ast(k_1)\;\varepsilon(p_2)\!\cdot\!\varepsilon^\ast(k_2)$\\[4pt] \noindent
     \end{minipage}
\\
\hline
   $B^\ast \bar K - B\bar K^\ast$ ($1^+$)
   & \begin{minipage}[c]{0.72\columnwidth}\centering
    \noindent\\[4pt]
    $V_{11}=-\displaystyle\frac{g^2}{2}\bigl(3D_{su}(m_\rho)+D_{su}(m_\omega)\bigr)\,\varepsilon(p_1)\!\cdot\!\varepsilon^\ast(k_1)$\\[4pt]
     $V_{12}=\displaystyle\frac{2g^2}{3}\bigl(-4D_t(m_\eta)+D_t(m_{\eta^\prime})-9D_\pi\bigr)\,k_1\!\cdot\!\varepsilon(p_1)\;p_2\!\cdot\!\varepsilon^\ast(k_2)$\\[4pt]
     $V_{21}=\displaystyle\frac{2g^2}{3}\bigl(-4D_t(m_\eta)+D_t(m_{\eta^\prime})-9D_\pi\bigr)\,k_2\!\cdot\!\varepsilon(p_2)\;p_1\!\cdot\!\varepsilon^\ast(k_1)$\\[4pt]
     $V_{22}=-\displaystyle\frac{g^2}{2}\bigl(3D_{su}(m_\rho)+D_{su}(m_\omega)\bigr)\,\varepsilon(p_2)\!\cdot\!\varepsilon^\ast(k_2)$\\[4pt] \noindent
     \end{minipage}
\\
\hline
 $ B^\ast\bar K^\ast$ ($1^+$)
 & \begin{minipage}[c]{0.72\columnwidth}\centering
     $\displaystyle\frac{g^2}{2}\bigl(3D_{su}(m_\rho)+D_{su}(m_\omega)\bigr)\,\varepsilon(p_1)\!\cdot\!\varepsilon^\ast(k_1)\;\varepsilon(p_2)\!\cdot\!\varepsilon^\ast(k_2)$
   \end{minipage}
\\[8pt]
\hline\hline
\end{tabular*}
\end{table}

In Table \ref{Tab:pot2} the functions $D(m)$ are defined as,
\begin{align}
D_{su}(m)=\frac{s-u}{t-m^2};\quad
D_{t}(m)=\frac{1}{t-m^2},
\end{align}
where $s=(p_1+p_2)^2,~t=(p_1-k_1)^2$, and $u=(p_1-k_2)^2$ are the Mandelstam variables. On the other hand, the pion propagator $D_{\pi}$ is give from {the time-order perturbation theory (TOPT)} \cite{Baru:2004kw,Lensky:2005hb}
\begin{align}
D_{\pi}
=\frac{1}{2E_{\pi}}\left[\frac{1}{\sqrt{s}-E_{p_2}-E_{k_1}-E_{\pi}+i\varepsilon}+\frac{1}{\sqrt{s}-E_{p_1}-E_{k_2}-E_{\pi}+i\varepsilon}\right],
\end{align}
where $E_{\pi}=\sqrt{m_{\pi}^2+(\vec{k}_1-\vec{p}_1)^2}$. The interaction given in Table~\ref{Tab:pot2} is projected in partial waves, see details in Appendix \ref{Sec:partial_wave}. See also Ref.~\cite{Shi:2024squ} for the partial wave projection of the $B\bar{K}^*-B^*\bar{K}$ system. {The $S$-wave interaction is attractive enough to generate states with spins $J^P=0^+,1^+,2^+$, while the contribution of the $D$-wave is found to be very small.}
The coupled-channel scattering amplitude is derived from the Lippmann-Schwinger (LS) equation,
\begin{align}
T_{ij}(E,p,k)=V_{ij}(E,p,k)+\sum_{k}\int_{\Lambda}\frac{d^3\vec q}{(2\pi)^3} V_{ik}(E,p,q)I_k(E,q)T_{kj}(E,q,k)\ ,
\label{Eq:T_matrix}
\end{align}
{where the indices $i,j,k$ refer to the channels considered. }
In the above equation, $I_k(E,q)$ is the usual Green's function,
\begin{align}
I_k(E,q)=\frac{\omega_1(q)+\omega_2(q)}{2\omega_1(q)\omega_2(q)}\frac{1}{E^2-\left[\omega_1(q)+\omega_2(q)\right]^2+i\epsilon},
\label{Eq:green}
\end{align}
where $w_{\alpha}(q)=\sqrt{m_{\alpha}^2+q^2}$. When treating the system with the unstable $\bar{K}^*$, its finite decay width should be incorporated, see details in Appendix~\ref{sect:kw}. In this case, the Green's function is modified to
\begin{eqnarray}
    I(E,q)=\frac{1}{4\,\omega_{\bar{K}^*}\omega_{B^{(*)}}(P^0-\omega_{\bar{K}^*}-\omega_{B^{(*)}}+i\frac{\Gamma}{2})},\label{eq:nl}
\end{eqnarray}
where $\Gamma$ denotes the decay width of the $\bar{K}^*$, which is taken as a momentum-dependent term,
\begin{align}
\Gamma\equiv\Gamma_{\bar K^*}(M_{\bar K^*})=\Gamma[{\bar K^*\to \bar K\pi}]\frac{m_{\bar K^*}^2}{M_{\bar K^*}^2}\left(\frac{p(M_{\bar K^*},m_{\bar K}, m_{\pi})}{p(m_{\bar K^*}, m_{\bar K}, m_{\pi}))}\right)^3\Theta(M_{\bar K^*}-m_{\bar K}-m_{\pi}).
\label{Eq:width_Kstar}
\end{align}
{Here the decay width of $\bar K^*$ associates with the imaginary part of the $\bar K \pi$ self-energy diagram, where the effective coupling $g$ for the $\bar K^*\to \bar K \pi$ decay is estimated from the decay width through the relation
\begin{align}
\Gamma[\bar K^*\to \bar K \pi]= \frac{g^2p(m_{\bar K^*}, m_{\bar K}, m_{\pi})^3}{24\pi m_{\bar K^*}^2}.
\end{align}
The real part of the self-energy diagram is absorbed by the bare $\bar K^*$ mass at the point $p=0$ to ensure that $m_{\bar K^*}$ represents the physical $\bar K^*$ mass.}
Its invariant mass is $M_{\bar K^*}^2=\left(E-\sqrt{m_{B^{(*)}}^2+q^2}\right)^2-q^2$ with $E=\sqrt{s}$. The Heaviside theta function is utilized to remove the effect of the anomalous threshold. The momentum in the center-of-mass frame is $p(M,m_1,m_2)=\sqrt{\lambda(M^2,m_{1}^2,m_{2}^2)}/2M$ with $\lambda$ the K\"all\'en function. For the $B^*\bar{K}^*$ loop function, we neglect the $B^*$ decay width as it is small compared to that of the $\bar{K}^*$ meson. 

In the complex $E$-plane, the second Riemann Sheet (RS) can be reached through the analytical continuation
\begin{align}
G^{\text{on},-}_k(E) = G^{\text{on},+}_k(E) + \frac{ip_k^{\text{on}}}{4\pi E}, 
\label{Eq:relation_RS}
\end{align}
where the $k$-channel on-shell Green's function in the physical RS is
\begin{align}
G^{\text{on},+}_k(E)&=\int\frac{d^3\vec q}{(2\pi)^3}I_k(E,q).
\end{align}
Considering the second RS of the $B^{(*)}\bar K \pi$ three-body cut, the momentum-dependent partial width can be replaced by
\begin{align}
\Gamma_i(M)\to -\Gamma_i(M)~~\text{for}~~\text{Im}(M)< 0,
\end{align}
where $\Gamma_i(M)$ represents partial width relevant to the three-body cut, and $M$ is invariant mass. See also for a discussion on the implementation of three-body effects Refs.~\cite{Doring:2009yv,Mai:2017vot,Sadasivan:2021emk,Du:2021zzh, Zhang:2024dth,Shi:2024squ}. {The pole positions of $B^{(*)}\bar K^{(*)}$ systems are determined on each RS by numerically solving the LS equation in Eq.~\eqref{Eq:T_matrix}. When the center-of-mass energy is above the corresponding $B^{(*)}\bar K \pi$ three-body threshold, we search for poles on the second RS associated with the relevant three-body cut. }
Finally, the couplings of a resonance $R$ to a two-body channel $i$ can be evaluated through the residue of the $T$-matrix, as follows,
\begin{align}
g_{i}g_{j}=\lim_{s\to E_R^2}(s-E_R^2)T_{ij}(E,p,k),
\label{Eq:coupling}
\end{align}
with $E_R$ the energy for the pole position of the resonance $R$.

\section{Numerical Result}\label{sec:result}

In the HGF and at LO for the meson-meson interaction (Weinberg-Tomozawa term), the only free parameter is the hard cutoff $\Lambda$, which regularizes the ultraviolet divergence in the Green's function of Eqs.~\eqref{Eq:green}. By interpreting $B_{sJ}(6063)^0$ as an isoscalar {structure near the $B\bar K^*$ threshold}, $\Lambda$ is determined from {a fit to the pole position within the experimental error bars of the mass and decay width}, using both, the on-shell and off-shell effective interactions. In the on-shell scheme, the fit yields $\Lambda=593\pm8$~MeV with $\chi^2/\text{d.o.f}=0.72$. Solving the integral LS equation in Eq.~\eqref{Eq:T_matrix}, the off-shell approach gives $\Lambda=775\pm8$ MeV with $\chi^2/\text{d.o.f}=1.15$. {If instead, we constrain $\Lambda$ using the pole position of $B_{sJ}(6114)^0$ as a resonance close to the $B^*\bar K^*$ threshold, we obtain similar values to the one above mentioned. One could think that the variation in the cutoff in the on-shell and off-shell approaches could lead to large uncertainties in the observables. However, this is not the case, since as one can see in Tables~\ref{Tab:onshpole_Ierr}-\ref{Tab:pole_IIerr} the differences between the pole positions in both approaches are indeed very small. When we estimate the systematic errors in the masses of the states obtained by conducting a simultaneous fit to the $B_{sJ}(6063)^0$ and $B_{sJ}(6114)^0$ masses and decay widths we obtain that the errors in the masses are about $7-8$~MeV, which are also quite small}.

\begin{table}[H]
\caption{{Predicted pole positions of the $B^{(*)}\bar K^{(*)}$ systems with the on-shell potentials. In the fourth column, ``$+$" and ``$-$" indicate that the poles lie on the physical and unphysical RS relative to the corresponding two-body cut, respectively. {For the $B^{(*)}\bar K^{*}$ systems, all poles lie on the second RS associated with the $B^{(*)}\bar K\pi$ three-body cut, which is not explicitly shown in the fourth column.} The quantity $g_{i}$ in the fifth column denotes the $S$-wave effective coupling. The sixth column, labeled ``Thr.", represents the relevant threshold. The mass splitting $E_b$ is defined as the difference between the real part of the pole position and the relevant threshold. We implement errors from $\Lambda=593\pm8$ MeV.}}
\label{Tab:onshpole_Ierr}
\renewcommand{\arraystretch}{1.2}
\begin{tabular*}{\columnwidth}{@{\extracolsep\fill}lcccccc}
\hline\hline 
 $J^{P}$ &  Channel & Pole [MeV] & RS  & $g_{i}$ [GeV] & Thr. [MeV] & $E_b$ [MeV]\\[3pt]     
\hline
$0^{+}$  & $ B\bar K$ & $5758.9\pm1.5$
     & $(+)$ & $23.0\pm0.5$ & $5775.0$ &$-16.1\pm1.5$\\[3pt]
\hline
$1^{+}$  & $ B^\ast\bar K$ & $5804.0\pm1.5$
     & $(+)$ & $23.0\pm0.5$ & $5820.0$ &$-16.0\pm1.5$\\[3pt]
\hline     
$1^{+}$  & $ B\bar K^\ast$ & $6109.0\pm1.4-(7.2\pm0.5)i$
     & $(+)$ & $39.5\pm0.5-(2.5\pm0.4)i$ & $6173.0$ &$-64.0\pm1.4$\\[3pt]
\hline
$0^{+}/1^+/2^+$  & $ B^\ast\bar K^\ast$ & $6154.1\pm1.5-(7.2\pm0.6)i$
     & $(+)$ & $38.8\pm0.4-(2.8\pm0.4)i$ &  $6218.0$ & $-63.9\pm1.5$\\[3pt]
\hline\hline
\end{tabular*}
\end{table}

In the on-shell approach, the coupled-channel effects between $B{}^*\bar K$ and $B\bar K{}^*$ {, and similarly, the $B\bar{K}$ and $B^*\bar{K}^*$ channels, are} neglected. {However, as it has already been mentioned, when the off-shell approach is accounted for, the pole positions obtained without neglecting this coupled-channel effect are very similar to the corresponding ones in the on-shell case, as we shall see. In addition, since the binding energy of the states obtained is much smaller than the threshold difference, this coupled-channel effect turns out to be very small.}
The resulting pole positions of isoscalar $B^{(*)}\bar K^{(*)}$ systems are listed in Table~\ref{Tab:onshpole_Ierr}. All poles are located on the first RS with respect to the corresponding two-body thresholds. The $B\bar K^*$ and $B^* \bar K^*$ molecules lie on the unphysical RS associated with the $B\bar K \pi$ and $B^*\bar K \pi$ three-body cuts, thereby preserving the analyticity of the $T$-matrix.
For the $B\bar K$ and $B^* \bar K$ systems, the binding energies are approximately 16~MeV, while for the $B\bar K^*$ and $B^* \bar K^*$ systems these are about $64$~MeV. The decay widths of the $B\bar K^*$ and $B^* \bar K^*$ molecules are nearly identical since the width of the $B^*$ is much smaller than that of the $\bar K^*$ and it can be neglected.\footnote{The dominant decay mode of $B^*$ is $B^*\to B\gamma$, with a decay width of only a few keV~\cite{Godfrey:2016nwn}. }
The deviation of the predicted widths from the experimental values is within $2\sigma$~\cite{LHCb:2020pet}. {The splitting of $S$-wave potentials for the $B^*\bar {K} ^*$ systems with different quantum numbers is suppressed by the bottom-meson mass, see Eq.~\eqref{Eq:epsilon_p1_k1}. As a result, the differences in the binding energies are not appreciable in Table~\ref{Tab:onshpole_Ierr} and the states are mostly degenerate for the $B^*\bar K^*$ molecules with $J^P=0^+$, $1^+$, and $2^+$.}

Taking into account the momentum dependence of the effective potential, the pole positions are obtained by solving the integral LS equation, as listed in Table~\ref{Tab:pole_IIerr}\footnote{The coupling for the $B^\ast B P$ vertex is taken as the
$D^*DP$ assuming heavy flavor symmetry~\cite{Gil-Dominguez:2024zmr}. Additionally, we checked that the pole positions obtained here are not very sensitive to this parameter.}. Compared with the results in Table~\ref{Tab:onshpole_Ierr}, the binding energies change only by a few MeV, and the decay widths of $B\bar K^*$ and $B^* \bar K^*$ molecules exhibit only minor variations. These results indicate that the on-shell approach provides a good approximation for describing the pole positions and effective couplings $g_i$'s. However, the corresponding hard cutoff differs significantly between the two schemes, implying that a common cutoff cannot be used to describe the $T$-matrix in both the on-shell and off-shell approaches.

{We predict shallow bound states in the $B\bar K$ and $B^* \bar K$ systems with binding energies of around $16$~MeV for the $J^P=0^+$ and $1^+$ states, respectively, using both, the on-shell and off-shell approaches, as listed in Tables~\ref{Tab:onshpole_Ierr} and \ref{Tab:pole_IIerr}. Our results are consistent with the theoretical predictions with respect to the Goldberger-Treiman relation~\cite{Bardeen:2003kt}, HMCHPT~\cite{Kolomeitsev:2003ac,Altenbuchinger:2013vwa}, and the previous LQCD result ~\cite{Gregory:2010gm}. We obtain smaller binding energies than the ones from the LQCD simulation including a base of approximately local-$q\bar{q}$ operators~\cite{Gayer:2024akw} of around $80-100$~MeV. The scarce lattice studies including also two-meson operators report slightly smaller binding energies (about $60$~MeV)~\cite{Lang:2015hza}. Previous unitarized chiral EFT approaches~\cite{Guo:2006rp,Guo:2006fu,Albaladejo:2016ztm,Albaladejo:2016lbb}, HGF~\cite{Sun:2018zqs}, and unquenched quark models~\cite{Yang:2022vdb,Ni:2023lvx} base their predictions for these states on calculations in the charm sector or the scarce lattice QCD simulations available in this sector, giving rise to binding energies in the range of $40-80$~MeV respect to the $B^{(*)}\bar{K}$ threshold. Instead, we have made predictions for these states which are based on the experimental information available in the $B_s$ sector in this work. The main difference comes here mostly from the value of the regulator, but we have also taken into account couple-channel effects and done the off-shell calculation for the first time, comparing with the results in the usual on-shell approach.   }

\begin{table}[H]
\caption{{Same as Table~\ref{Tab:onshpole_Ierr}, but using the off-shell interaction. The $g_i$ in the fifth column is the S-wave effective coupling, except for the $J^P=2^+$ case, where we also include the $D$-wave coupling. Since the $D$-wave coupling is generally too small, we do not show the numerical result in the table for other cases. The errors shown in the table are evaluated from the systematic error in the cutoff parameter, $\Lambda=775\pm8$ MeV.}}
\label{Tab:pole_IIerr}
\renewcommand{\arraystretch}{1.0}
\begin{tabular*}{\columnwidth}{@{\extracolsep\fill}lcccccc}
\hline\hline 
 $J^{P}$ &  Channel & Pole [MeV] & RS  & $g_{i}$ [GeV] (wave) & Thr. [MeV] & $E_b$ [MeV]\\[3pt]     
\hline
\multirow{4}{*}{$0^{+}$}  & \multirow{4}{*}{$B\bar K$--$B^\ast\bar K^\ast$} & \multirow{2}{*}{$5752.4\pm1.3$}
     & \multirow{2}{*}{$(+,+)$} & $24.9\pm0.4$ & \multirow{2}{*}{$5775.0$} & \multirow{2}{*}{$-22.6\pm1.3$}\\[3pt]
&&&& $-5.9\pm0.3$ & &\\
  & & \multirow{2}{*}{$6153.8\pm1.5-(6.2\pm0.3)i$}
     & \multirow{2}{*}{$(-,+)$} & $0.8\pm0.2+(1.1\pm0.3)i$ & \multirow{2}{*}{$6218.0$} & \multirow{2}{*}{$-64.2\pm1.5$}\\[3pt]
&&&& $-45.1\pm0.6+(2.4\pm0.4)i$ & &\\
\hline
\multirow{4}{*}{$1^{+}$}  & \multirow{4}{*}{$B^\ast\bar K$--$B\bar K^\ast$} & \multirow{2}{*}{$5800.0\pm0.8$}
     & \multirow{2}{*}{$(+,+)$} & $25.1\pm0.4$ & \multirow{2}{*}{$5820.0$} & \multirow{2}{*}{$-20.0\pm0.8$}\\[3pt]
&&&& $-4.8\pm0.3$ & &\\
  & & \multirow{2}{*}{$6108.0\pm1.6-(10.3\pm0.4)i$}
     & \multirow{2}{*}{$(-,+)$} & $0.53\pm0.12+(1.2\pm0.3)i$ & \multirow{2}{*}{$6173.0$} & \multirow{2}{*}{$-65.0\pm1.6$}\\[3pt]
&&&& $-52.5\pm0.6+(1.5\pm0.3)i$ & &\\
\hline
\multirow{1}{*}{$1^+$}  & \multirow{1}{*}{$B^\ast\bar K^\ast$} & \multirow{1}{*}{$6155.7\pm1.7-(6.4\pm0.4)i$}
     & \multirow{1}{*}{$(+)$} & $45.1\pm0.5-(2.2\pm0.2)i$ &  \multirow{1}{*}{$6218.0$} & \multirow{1}{*}{$-62.3\pm1.7$}\\[3pt]
\hline
\multirow{2}{*}{$2^{+}$}  & \multirow{2}{*}{$B\bar K$--$B^\ast\bar K^\ast$} & \multirow{2}{*}{$6154.0\pm1.4-(6.2\pm0.3)i$}
     & \multirow{2}{*}{$(-,+)$} & $\left[69\pm4+(2.3\pm0.3)i\right]\times10^{-3}$ & \multirow{2}{*}{$6218.0$} & \multirow{2}{*}{$-64.0\pm1.4$}\\[3pt]
&&&& $-45.1\pm0.6+(2.4\pm0.4)i$ & &\\
\hline\hline
\end{tabular*}
\end{table}

Compared to the heavy quark partners, $D_{s0}(2317)$ and $D_{s1}(2460)$, the $B\bar K$ and $B^*\bar K$
 molecules are expected to exhibit larger relative momenta in their decay channels. Notably, the relative momenta of $B_s\pi$ and $B_s^* \pi$ systems, evaluated at the masses of the predicted states, are about $350$~MeV. Therefore, it is natural to suggest searching for the $B\bar K$ and $B^*\bar K$ molecules in the $B_s\pi$ and $B_s^* \pi$ invariant mass distributions.

\section{Summary}\label{sec:sum}

Based on the HQS, the hadronic molecular interpretation of the states $D_{s0}(2317)$ and $D_{s1}(2460)$ suggests the existence of their bottom counterparts as $B\bar K$ and $B^* \bar K$ molecules. In this work,
we have derived the effective potential within the HGF by considering the exchange of light vector and pseudoscalar mesons. Notably, we take into account the three-body effect in the $B^*\bar K$--$B\bar K^{*}$ and $B^* \bar K^{*}$ systems. This originates from the self-energy contributions of the unstable $\bar K^*$ meson, as well as from the one-pion-exchange potential in the $B^*\bar K$--$B\bar K^{*}$ coupled-channel system.
The properties of the $B^{(*)}\bar K^{(*)}$ molecules are then investigated by solving the LS equation using both the on-shell and off-shell approaches. In our analysis, we only have one free parameter, the hard cutoff $\Lambda$, which is determined by reproducing the pole position of $B_{sJ}(6063)^0$~\cite{LHCb:2020pet}, interpreting it as a $B\bar K^*$ molecule. {The splitting of about $50$~MeV obtained here between the $J^P=1^+$ and $2^+$ states as $B\bar K^*$($B^*\bar{K}^*$) molecules agrees well with the experimental one related to the $B_{sJ}(6063)^0$ and the $B_{sJ}(6114)^0$ states. }

Our results, {based on the experimental information in the $B_s$ sector}, support the existence of six $B^{(*)}\bar K^{(*)}$ molecules. Near the $B\bar K$ and $B^{*} \bar K$ thresholds, two bound states with quantum numbers $J^P=0^+$ and $1^+$, respectively, are obtained {with a binding energy of around $16$~MeV}. 
A pole is also found close to the $B\bar K^*$ threshold with a binding energy of about 64 MeV, which possibly corresponds to $B_{sJ}(6063)^0$. In the $B^*\bar K^*$ system, the states are predicted with a binding energy of approximately $64$~MeV for $J^P=0^+$,$1^+$, and $2^+$. Among them, the $1^+$ $B^*\bar K^*$ molecule may be related to the $B_{sJ}(6114)^0$. 
Furthermore, we find that the on-shell approach describes the low-energy scattering of hadron pairs as well as the off-shell approach, provided that the cutoff is tuned.

\begin{acknowledgements}

R.~Molina acknowledges support from the ESGENT program with Ref. ESGENT/018/2024 and the PROMETEU program with Ref. CIPROM/2023/59, of the Generalitat Valenciana, and also from the Spanish Ministerio de Economia y Competitividad and European Union (NextGenerationEU/PRTR) by the grant with Ref. CNS2022-136146, Ref. PID2023-147458NB-C21, and CEX2023-001292-S. This project has received funding from the European Union Horizon 2020 research and innovation program under the program H2020-INFRAIA-2018-1, grant agreement No. 824093 of the STRONG-2020 project.

\end{acknowledgements}

\appendix
\section{Loop integral with $\bar{K}^\ast$ decay width}\label{sect:kw}
\begin{figure}[tbh]   \includegraphics[height=0.3\columnwidth]{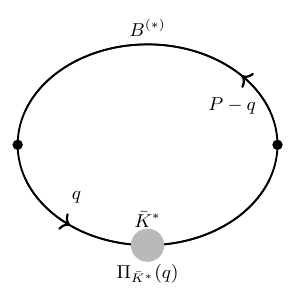}
   \caption{Feynman diagram for the loop interaction of the $ B^{(\ast)} \bar K^\ast$ system with the self-energy of $K^\ast$.}
    \label{Senergy}
\end{figure}
In order to include the $\bar{K}^*$ decay width in the $B^{(\ast)}\bar{K}^*$ loop, we consider the diagram of the loop function in Fig. \ref{Senergy} modified by the imaginary part of the self-energy of the $\bar{K}^*$. Starting from the usual formula of the loop integral,
\begin{eqnarray}
    G(P)=i\int\frac{d^4 q}{(2\pi)^4} \frac{1}{q^2-m_1^2+i\epsilon}\frac{1}{(P-q)^2-m_2^2+i\epsilon}.
\end{eqnarray}
with $P^0=E$, and $\vec{P}=0$ in the center of mass frame. Since in this case we are dealing with heavy mesons and strange vector mesons, it is sufficient to keep only the positive energy part of the propagators. 
By inserting the self-energy diagram of the $\bar K^*$, the $G(E)$ is reduced to
\begin{eqnarray}
    G(P^0)\simeq i\int\frac{d^4\,q}{(2\pi)^4}\frac{1}{4\,\omega_{K^*}\omega_{B^{(\ast)}}}\frac{1}{q^0-\omega_{K^*}+i\frac{\Gamma}{2}}\frac{1}{P^0-q^0-\omega_{B^{(\ast)}}+i\epsilon}\ ,
\end{eqnarray}
where $\omega_{K^*}=\sqrt{\vec{q}\,^2+m_{K^*}^2}$, $\omega_{B^{(\ast)}}=\sqrt{\vec{q}\,^2+m_{B^{(\ast)}}^2}$. We integrate the $q^0$-component using the residue theorem and obtain the Green's function
\begin{eqnarray}
    G(P^0)=\int \frac{d\,q}{8\pi^2} \frac{ \vec{q}\,^2}{\omega_{K^*}\omega_{B^{(\ast)}}(P^0-\omega_{K^*}-\omega_{B^{(\ast)}}+i\frac{\Gamma}{2})}\ .
\end{eqnarray}

\section{Partial wave projection}\label{Sec:partial_wave}

We define state in the basis of orbital angular momentum $l$ and its projection $m$ as
\begin{align}
\left|l m, s_1 s_2\right\rangle\equiv\frac{1}{\sqrt{4 \pi}} \int d \hat{p}\, Y_{l}^m(\hat{p})\left|\vec{p}, s_1 s_2\right\rangle,
\end{align}
where $s_1$ and $s_2$ are the spins of the two particles, $\vec{p}$ is the center-of-mass (c.o.m.) three-momentum, and $\hat{p}=\vec{p}/|\vec{p}|$.
In the spin-orbit basis, the two-body state is written as
\begin{align}
 \left|LS, J M\right\rangle=\frac{1}{\sqrt{4 \pi}} \sum_{\substack{\sigma_1, \sigma_2 \\
\mu, m }} \int d \hat{p}\, Y_{L}^m(\hat{p})\left(\sigma_1 \sigma_2 \mu \mid s_1 s_2 S\right)(m \mu M \mid L S J) \left|\vec{p}, \sigma_1 \sigma_2\right\rangle,
\end{align}
where $L$ and $S$ are the orbital angular momentum and total spin of the two-body system, $J$ and $M$ denote the total angular momentum and its third component, and $(m_1m_2m_3|j_1j_2j_3)$ stands for the Clebsch--Gordan coefficient with $m_i$ being the third component related to $j_i$.

The partial-wave amplitude for the two-body scattering  is~\cite{Gulmez:2016scm,Oller:2019rej}
\begin{align}
T_{L S ; \bar{L} \bar{S}}^{J}=&\frac{Y_{\bar{L}}^0(\hat{z})}{2 J+1} \sum_{\substack{\sigma_1, \sigma_2, \bar{\sigma}_1 \\
\bar{\sigma}_2, m}} \int d \hat{p}^{\prime \prime} Y_{L}^m\left(\hat{p}^{\prime \prime}\right)^*\left(\sigma_1 \sigma_2 M \mid s_1 s_2 S\right)(m M \bar{M} \mid L S J)\left(\bar{\sigma}_1 \bar{\sigma}_2 \bar{M} \mid \bar{s}_1 \bar{s}_2 \bar{S}\right)\nonumber\\
&\times(0 \bar{M} \bar{M} \mid \bar{L} \bar{S} J)
\left\langle\vec{p}^{\,\prime \prime}, \sigma_1 \sigma_2|\hat{T}||\vec{p}| \hat{z}, \bar{\sigma}_1 \bar{\sigma}_2\right\rangle,
\label{pwaveeq}
\end{align}
where $\vec{p}$ ($\vec{p}^{\,\prime\prime}$), $\bar{L}$ ($L$), and $\bar{S}$ ($S$) refer to the c.o.m. three-momentum, orbital angular momentum, and spin of the initial (final) state, respectively. Here the initial c.o.m. momentum is chosen along the $z$ axis.
The $S$-wave components of the potential $V=D(s,t,u)\varepsilon(p_1)\cdot\varepsilon^*(k_1)\varepsilon(p_2)\cdot\varepsilon^*(k_2)$ are reduced to
\begin{align}
V_0^{SS}(p,k)&=\int_{-1}^{1}dz~D(s,t,u)\left[\frac{1}{2}-\frac{pkz}{6}\left(\frac{1}{m_{k_1}m_{p_1}}+\frac{1}{m_{k_2}m_{p_2}}\right)+\frac{p^2k^2}{6m_{p_1}m_{k_1}m_{p_2}m_{k_2}}\right],\nonumber\\
V_1^{SS}(p,k)&=\int_{-1}^{1}dz~D(s,t,u)\left[\frac{1}{2}-\frac{pkz}{6}\left(\frac{1}{m_{k_1}m_{p_1}}+\frac{1}{m_{k_2}m_{p_2}}\right)\right],\nonumber\\
V_2^{SS}(p,k)&=\int_{-1}^{1}dz~D(s,t,u)\left[\frac{1}{2}-\frac{pkz}{6}\left(\frac{1}{m_{k_1}m_{p_1}}+\frac{1}{m_{k_2}m_{p_2}}\right)+\frac{p^2k^2}{30m_{p_1}m_{k_1}m_{p_2}m_{k_2}}(3z^2-1)\right],
\label{Eq:epsilon_p1_k1}
\end{align}
where the subscript $J$ in $V_J$ denotes the total angular momentum of the two-body system. In the above reduction, we adopt the approximation $E_{i}/m_{i}\simeq 1$, with the three-momenta $p=|\vec p_1|$ and $k=|\vec k_1|$ in the c.o.m frame. The variable $z=\cos\theta$, where $\theta$ is the angle between the momenta $p_1$ and $k_1$.

The $S$-wave components of the potential $V=D(s,t,u)p_1\cdot\varepsilon^*(k_1)p_2\cdot\varepsilon^*(k_2)$ are reduced to
\begin{align}
V_0^{SS}(p,k)&=\int_{-1}^{1}dz~D(s,t,u)\left[\frac{k^2m_{p_1}m_{p_2}}{2\sqrt{3}m_{k_1}m_{k_2}}+\frac{p^2}{2\sqrt{3}}-\frac{pkz}{2\sqrt{3}}\left(\frac{m_{p_1}}{m_{k_1}}+\frac{m_{p_2}}{m_{k_2}}\right)\right],\nonumber\\
V_2^{SS}(p,k)&=0.
\label{Eq:epsilon_p1_p2}
\end{align}
The potential with polarization $V=D(s,t,u)k_1\cdot\varepsilon(p_1)k_2\cdot\varepsilon(p_2)$ leads to the same form of partial-wave potential as in Eq. \ref{Eq:epsilon_p1_p2}, but with masses of the initial and final states exchanged, i.e. $m_{p_1(p_2)}\rightarrow m_{p_2(p_1)}$ and $m_{k_1(k_2)}\rightarrow m_{k_2(k_1)}$. 

\bibliography{ref.bib}
\end{document}